\documentclass[a4paper]{jpconf}
\usepackage{graphicx}
\usepackage{color}
\usepackage{subfigure}
\usepackage{units}
\usepackage{booktabs}
\usepackage{amsmath}
\usepackage{amssymb}
\usepackage{feynmf}
\usepackage{textfit}
\usepackage{textcomp}
\usepackage[abs]{overpic}
\usepackage{ragged2e}
\usepackage{hyperref} 
\usepackage{tikz}
\begin{document}
\title{Bound-free pair production in heavy-ion collisions at high energies}

\author{Rainer Schicker}

\address{Physikalisches Institut, Im Neuenheimer Feld 226, 69120 Heidelberg}

\ead{schicker@physi.uni-heidelberg.de}

\begin{abstract}

The electromagnetic process of bound-free electron pair production in 
heavy-ion collisions at high energies is reviewed. The importance of this 
process for producing secondary beams is outlined. Single free electron pair 
production is presented, and the bound-free pair production process is 
introduced.  Double pair production is discussed, and an estimate of the 
bound-free pair constrained photon-photon luminosity is given.

\end{abstract}

\section{Introduction}

The process of bound-free pair production in heavy-ion collisions at high 
energies is of interest both from an experimental and a theoretical view point.
From an experimental view point, bound-free pair production is of relevance 
since it is an important contribution to the life time of a heavy-ion beam at 
high energies, together with single and multiple neutron emission following 
Coulomb excitation due to absorption of one or several photons\cite{Bruce}. 
The bound-free pair process changes the electric charge state 
of beam particles by one unit, and a secondary beam of different magnetic 
rigidity is hence emerging from the target position. This secondary beam 
is transported by the magnets of the beam system, and will eventually hit 
the vacuum pipe somewhere downstream. Depending on beam species, beam energy 
and beam intensity, the local deposition of energy at the point of impact 
might do harm to the vacuum and/or magnet infrastructure, and therefore 
needs to be studied carefully. The installation of beam collimators at or 
before the point of impact might be necessary to absorb the energy of this 
\mbox{secondary beam \cite{Jowett}.} Detectors, integrated into the 
collimators, and capable of identifying the secondary beam particles on an 
event-by-event basis, would give access to interesting measurements such as 
single and double bound-free pair production cross sections.    

The process of free and bound-free electron pair production is of 
electromagnetic origin, and is hence in principle calculable within the 
framework of QED. The large electric charge of a heavy ion leads, however, to 
a large electromagnetic coupling at the ion-photon vertex, and the summation 
of many higher order terms in the perturbation series is needed for getting 
accurate results. For example, the single free pair production cross section at 
Born level is modified by substantial Coulomb corrections on the
order of 14\% at LHC energies \cite{Ccorr}. Multiple pair 
processes, for example production of a single free electron pair in 
conjunction with a bound-free lepton pair, are also subject to unitarity  
and Coulomb corrections.  

\section{Secondary Pb-beams at hadron colliders}

Electromagnetic processes can result in secondary ion beams originating
from the target position. Here, secondary beam refers to beam particles
which possess a magnetic rigidity different than the primary beam. 

\begin{figure}[h]
\begin{center}
\begin{overpic}[width=.8\textwidth]{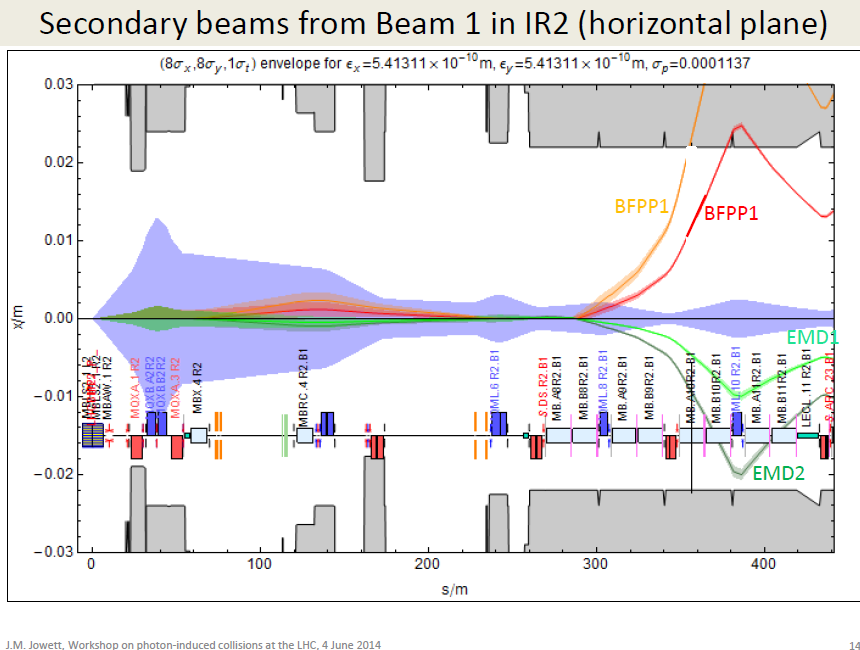}
\put(126.,81.){\textcolor{orange}{\footnotesize $Pb^{80+}_{208}$}}
\put(126.,64.){\textcolor{red}{\footnotesize $Pb^{81+}_{208}$}}
\put(126.,44.){\textcolor{green}{\footnotesize $Pb^{82+}_{207}$}}
\put(126.,38.){\textcolor{gray}{\footnotesize $Pb^{82+}_{206}$}}
\thicklines
\color{blue}
\put(78.4,9.0){\vector(2,0) {46.}}
\put(78.4,4.0){\textcolor{blue}{\it cold section: LHC dipoles}}
\color{black}
\end{overpic}
\end{center}
\caption{Beam transport calculation of Pb-beams at IR2 of the LHC, 
taken from Ref. \cite{JJ}}
\label{fig1}
\end{figure}

In Figure \ref{fig1}, the transport of the primary and secondary Pb-beams 
originating from an interaction region (IR) at the LHC is shown. The beams are 
shown here for the case of IR2, however similar conditions exist at the other 
IRs. The horizontal scale is shown in units of meters, with the target position
located at the origin. The grey shaded boxes at the top and bottom of the plot 
denote the different magnets of the beam transport system. The cold section of 
the LHC dipoles starts at a distance of about 260 m from the IR, and is 
indicated in \mbox{Figure \ref{fig1}} by a continous grey shaded area. On the 
vertical axis, the relative location of the beam and its size are indicated. 
Here, relative denotes the distance to the beam of nominal 
\mbox{momentum $p_{0}$} traveling on axis of the magnet beam transport system.
The 8$\sigma$ envelope of the primary Pb-beam is indicated by the blue 
contour. The secondary beam  Pb$_{208}^{81+}$ generated by single bound-free pair 
production is shown in Figure \ref{fig1} by the red line labeled BFPP1,
whereas the secondary beam  Pb$_{208}^{80+}$ generated by double bound-free pair 
production is shown by the orange line labeled BFPP1.
Similarly, the secondary Pb-beams Pb$_{207}^{82+}$ and Pb$_{206}^{82+}$ resulting 
from one and two neutron emission following Coulomb excitation are indicated 
by  the green EMD1 line and the grey EMD2 line, repectively. These secondary 
beams will hit the beam pipe at the location of the third or fourth dipole,
i.e. in the cold section. At the LHC, there is no space for additional detectors
at that location for measuring these secondary beams. Discussions have, 
however, started for the next generation hadron collider, the Future Circular 
Collider(FCC), where integration of such detector systems might be feasible
if taken into consideration at the early design stage of the machine lattice. 

\section{Single and multiple electron pair production}

The strong electromagnetic field of heavy-ion beams results in a variety of 
two-photon processes. In the Equivalent Photon Approximation (EPA), this 
electromagnetic field is expressed by the action of an ensemble of equivalent 
photons \cite{Fermi,Williams,Weizs,Budnev}. The cross section of heavy-ion 
induced photon-photon processes can be expressed as 
\vspace{-0.2cm}
\begin{eqnarray}
\sigma^{EPA}_{PbPb \rightarrow PbPb\,X} = \int\!\!\int dn_{1,\gamma} \; dn_{2,\gamma} \; 
\sigma_{\gamma\gamma \rightarrow x}(\omega_1\,\omega_2)  
\label{eq:ggcross}
\end{eqnarray}
with dn$_{i,\gamma}$ the photon flux of beam i, and $\sigma_{\gamma\gamma}$ the 
elementary \mbox{cross section of the process under study.}
\vspace{-.6cm} 
\begin{figure}[h]
\begin{center}
\begin{minipage}[h]{0.32\textwidth}
\begin{overpic}[width=.9\textwidth]{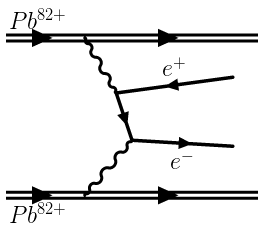}
\end{overpic}
\end{minipage}
\hspace{1.2cm}
\begin{minipage}[h]{0.26\textwidth}
\begin{overpic}[width=.9\textwidth]{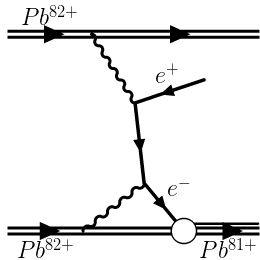}
\end{overpic}
\end{minipage}
\end{center}
\vspace{-.4cm}
\caption{Free electron pair production on the left, 
bound-free pair production on the right.}
\label{fig2}
\end{figure}

\vspace{-.2cm}
In Figure \ref{fig2} on the left, the Born level diagram for free electron 
pair production is shown. The cross section of this process  is given at 
tree level according to Eq. \ref{eq:ggcross} by integration of the product of 
differential photon fluxes with the elementary cross section $\gamma\gamma 
\rightarrow e^{+}e^{-}$. 
At LHC energies, the Born level cross section for free electron pair
production in the Pb-Pb system is approximately $\sim$ 200 kb.
In Figure \ref{fig2} on the right, the Born level diagram for bound-free
pair production is shown. Here, the electron gets bound by the Coulomb 
potential of one of the Pb-nuclei, resulting in a cross section for production
of a Pb-nucleus of charge state Pb$^{81+}$ of:
\begin{eqnarray}
\sigma^{EPA}_{(PbPb \rightarrow PbPb^{81+}\,+\,e^{+})} = \int dn_{1,\gamma} \; 
\sigma_{(\gamma\,+\,Pb^{82+}\rightarrow Pb^{81+}\,+\,e^{+})}(\omega_1).  
\label{eq:bfreecross}
\end{eqnarray}

The cross section for bound-free pair production is calculated according 
to Eq. \ref{eq:bfreecross} by folding the photon flux with the 
cross section of the elementary process 
\mbox{$\gamma + Pb^{82+} \rightarrow Pb^{81+} + e^{+}$ \cite{Agger}.} 
This cross section is approximately 270 b per beam for the Pb-Pb system 
at LHC energies\cite{bfcross}.
Cross section studies differential in the transverse momentum $p_{T}$
of the emitted positron have been discussed in Ref. \cite{Serbo1}. 

\vspace{-.3cm} 
\begin{figure}[h]
\begin{center}
\begin{minipage}[h]{0.30\textwidth}
\begin{overpic}[width=.9\textwidth]{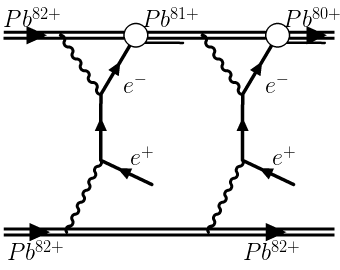}
\end{overpic}
\end{minipage}
\hspace{.2cm}
\begin{minipage}[h]{0.30\textwidth}
\begin{overpic}[width=.9\textwidth]{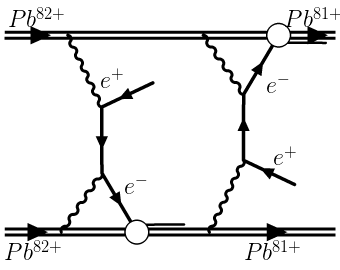}
\end{overpic}
\end{minipage}
\hspace{.2cm}
\begin{minipage}[h]{0.30\textwidth}
\begin{overpic}[width=.9\textwidth]{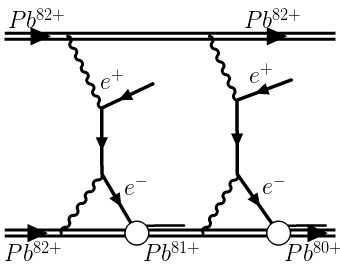}
\end{overpic}
\end{minipage}
\end{center}
\vspace{-.4cm}
\caption{Double bound-free pair production: double production on upper 
beam (left), single production on both upper and lower beam (middle),
double production on lower beam (right).}
\label{fig3}
\end{figure}

\vspace{-0.2cm}
Double bound-free pair production can occur in different configurations. 
The bound-free single pair production can happen twice on the same Pb-nucleus
as shown on the left and right of Figure \ref{fig3}. Here, the final state is
composed of the system  (Pb$^{82+}$ + Pb$^{80+}$). Double production can also
take place as single bound-free pair production on both of the two Pb-nuclei, 
leading to a final state (Pb$^{81+}$\:+\:Pb$^{81+}$) as shown in the middle of 
Figure \ref{fig3}. The cross section of this process is 
$\sigma_{2xBFPP}$(PbPb,LHC) $\sim$ 11-12 mb \cite{Serbo2}.

\section{Tag on electromagnetic processes}

Double production of electron pairs can also take place in mixed form, 
i.e. production of a free pair in conjunction with a  bound-free pair. This 
process is of particular interest since the forward scattered Pb-nucleus 
of different charge state carries the tag of an electromagnetic process. 
This tag can serve a two-fold purpose. First, this tag can be used for the 
measurement of the single and double pair production cross section. As shown in 
Figure \ref{fig1}, the secondary beams Pb$^{81+}$ and  Pb$^{80+}$ are separated 
by a transverse distance of approximately one centimeter at the point of 
impact, and hence can be identified by detectors with position resolution of 
order of millimeter. Additionally, the signal from these events can be used 
as trigger for readout of the midrapidity detectors in the search for triggered
two-photon events at midrapidity. Second, if the detector readout is triggered
from activity within the midrapidity detectors, or for detector systems with 
continous readout, the tag from the forward secondary beams can be used
in the analysis of coincidences between forward and midrapidity activity.  

\begin{figure}[h]
\begin{center}
\begin{minipage}[h]{0.32\textwidth}
\begin{overpic}[width=.9\textwidth]{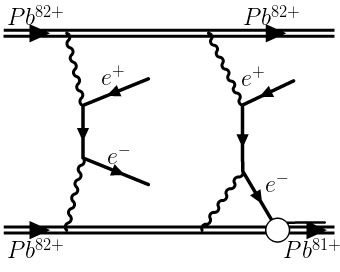}
\end{overpic}
\end{minipage}
\hspace{1.2cm}
\begin{minipage}[h]{0.32\textwidth}
\begin{overpic}[width=.9\textwidth]{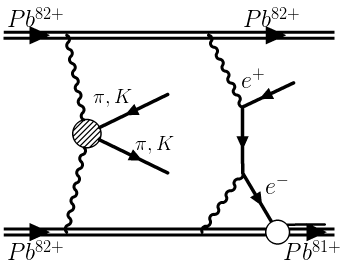}
\end{overpic}
\end{minipage}
\end{center}
\caption{Double  production of free and bound-free electron pair on the left, 
bound-free electron pair and free pion/kaon pair on the right.}
\label{fig4}
\end{figure}

The double production of a free and a bound-free electron pair is shown 
in Figure \ref{fig4} on the left. On the right side, double production
of a free pion/kaon pair and a bound-free electron pair is displayed.
Common to these two different processes is the forward scattered Pb$^{81+}$.
The detection of this forward Pb$^{81+}$-nucleus with coincident measurement
of central activity hence allows to select two-photon production
processes with very small background, an otherwise very difficult
endeavour at hadron colliders at high energies.

\section{Bound-free pair constrained photon-photon luminosity}

A constrained photon luminosity can be associated to each of the forward 
scattered secondary Pb-beams. This constrained photon luminosity represents the 
electromagnetic field of the secondary Pb-beam, and expresses the possibility
that this electromagnetic field can generate two-photon events by interacting
with the electromagnetic field of a nucleus of the other beam. This constrained
photon luminosity is dominated by the Pb$^{81+}$ beam since the double pair 
cross section is reduced by at least 4 orders of magnitude as compared to the 
single pair cross section.

An estimate of this constrained photon luminosity is given below 
for the case of muon pair production. The differential photon luminosity
for free muon pair production is given as 

\begin{eqnarray}
\frac{d^{2}L^{\it (st)}}{d\omega_1d\omega_2}=\! 
\frac{d\sigma_{AA \rightarrow AA\mu\mu}}{d\omega_1d\omega_2 
\sigma_{\gamma\gamma \rightarrow \mu\mu}}=\! 
\frac{dn(\omega_1)}{d\omega_1} \frac{dn(\omega_2)}{d\omega_2}=\!
\frac{4Z^{4}\alpha^{2}}{\pi^{2}\omega_1\omega_2} 
{\text{log}}\frac{\gamma}{R\omega_1} 
{\text{log}}\frac{\gamma}{R\omega_2}.
\label{eq:muonfree}
\end{eqnarray}

The differential luminosity for production of free muon pair plus
bound-free electron pair can be expressed as

\vspace{-.2cm}
\begin{eqnarray}
 \frac{d^{2}L^{bfree}}{d\omega_1d\omega_2}=\! 
\frac{dP_{\mu\mu}(b)}{d\omega_1d\omega_2 
\sigma_{\gamma\gamma \rightarrow \mu\mu}} P_{ee}(b,Z_1,Z_2) d^{2}b=\! 
\frac{Z^{4}\alpha^{2}}{\pi^{4}\omega_1\omega_2} F(b) P_{ee}(b,Z_1,Z_2) d^{2}b, 
\label{eq:muonbfree}
\end{eqnarray}

with the function F(b) defined by

\vspace{-.2cm}
\begin{eqnarray}
F(b) = \int_{R}^{\infty} \frac{d^2b_1 d^2b_2}{b^2_1b^2_1} 
\delta({\bf b_1-b_2-b}) \approx \frac{4\pi}{b^2} {\text{log}}\frac{b}{2R}.
\label{eq:fb}
\end{eqnarray}

In the integration over b in Eq. \ref{eq:fb}, the main contribution
comes \mbox{from the range $ 2R < b < 1/m_{e}$,} where   
$P_{ee}(b,Z_1,Z_2) \sim A = 1.65 \cdot 10^{-3}$ \cite{Serbo2}.

Substitung into Eq. \ref{eq:muonbfree} yields

\begin{eqnarray} 
\frac{d^{2}L^{bfree}}{d\omega_1d\omega_2}=\! A 
\frac{4Z^{4}\alpha^{2}}{\pi^{2}\omega_1\omega_2} \Big[{\text{log}} 
\frac{1}{2Rm_e}\Big]^2, 
\hspace{1.cm}  
\frac{d^{2}
L^{bfree}}{d\omega_1d\omega_2}
\sim \mathcal{O}(10^{-3}) \frac{d^{2}L^{st}}{d\omega_1d\omega_2},
\label{eq:muonbfree2}
\end{eqnarray}

hence the bound-free constrained photon luminosity is reduced at order 
$\mathcal{O}(10^{-3})$ as compared to the standard photon luminosity.

\section{Discussion}

The possibility of measuring secondary Pb-beams generated by bound-free 
electron pair production in coincidence with additional two-photon processes 
addresses a list of physics issues.

\begin{itemize} 

\item The cross section for bound-free pair production in heavy-ion collisions
can be calculated at tree level according to Eq. \ref{eq:bfreecross}. 
This cross section is modified by higher order terms due to unitarity and 
Coulomb corrections.

\item Double bound-free pair production is possible in different
configurations as discussed above. The measurement of the cross section of the
(Pb$^{82+}\,+\,$Pb$^{80+}$) and (Pb$^{81+}\,+\,$Pb$^{81+}$) final state permits to
test whether $\sigma$(Pb$^{81+}\,+\,$Pb$^{81+}$) $\sim$ 2 x 
$\sigma$(Pb$^{82+}\,+\,$Pb$^{80+}$) as derived in the impact parameter
representation \cite{Serbo2}. The measurement of these two cross sections 
allows to test the contributions of unitarity and Coulomb 
corrections to these two different final states.

\item Multiple production of N lepton pairs can be of mixed form, i.e. 
bound-free pair production with production of N-1 free lepton pairs\cite{Baur}.

\item Double pair production can be of mixed form, i.e. bound-free 
pair production in association with a pion or kaon pair.

\item Multiple interactions can be of mixed form, i.e. bound-free 
pair production in association with photoproduction of a resonance, for 
example $\eta,\eta^{'}$.

\item Multiple interactions can be of mixed form, i.e. bound-free pair 
production in association with two-photon processes beyond the resonance
region.
 
\item Multiple interactions can be of mixed form, i.e. bound-free
pair production in association with two-photon processes 
$\gamma\gamma \rightarrow \gamma\gamma$. 
This two-photon scattering receives contributions from lepton loops,
quark loops, and, possibly, the monopole loop 
\cite{Ginzburg},\cite{L3_mono},\cite{D0_mono}.

\item Multiple production of lepton pairs can be accompanied by internal
excitations of the nucleus, either in an intermediate state between two 
photon exchanges, or in the final state. The contribution of virtual internal
excitations to the double bound-free cross section is 
estimated to be $<$\,25\% which is well within the accuracy
of the \mbox{present calculations \cite{Serbo2}.}
Excitations of the final state Pb$^{81+}$ result with high probability in 
the emission of one or several neutrons, and lead to the generation of 
additional secondary beams. The experimental detection of hybrid beams 
such as Pb$_{207}^{81+}$ and Pb$_{206}^{81+}$ depends on the beam optics, and 
on the placement  of the detectors used  for secondary beam detection. 

\end{itemize} 

The approach of measuring these reaction channels in coincidence
with the bound-free produced Pb$^{81+}$ is expected to result
in a much improved signal to background ratio. In this approach, the 
coincident count rate is, however, reduced by the factor
of bound-free constrained to standard photon luminosity. As derived
above, this reduction factor is, for example, estimated to be on the
order $\mathcal{O}(10^{-3})$ for free muon pair production. 
This reduction needs to be evaluated for each measurement above 
individually due to the different cross sections involved.

\section{Summary and outlook}

The process of single and double-bound free pair production
in heavy-ion collisions at high energies is reviewed.
Double pair production in mixed form of free pair and bound-free pair 
is discussed. The experimental constraints of such measurements
are summarised.

\section{Acknowledgements}

The author gratefully acknowledges fruitful discussions with Valeriy 
Serbo, Otto Nachtmann and John Jowett on issues presented in this study.
This work is supported by the German Federal Ministry of Education and 
Research under promotional reference 05P15VHCA1.


\section*{References}
\begin{thebibliography}{9}

\bibitem{Bruce} R. Bruce, S. Gilardoni, J. Jowett and D. Bocian,
Phys.Rev.ST Accel.Beams 12 (2009) 071002, arXiv:0908:2527. 

\bibitem{Jowett} J. Jowett, J.Phys.Conf.Ser.312 (2011) 102017, arXiv:1109.0135.

\bibitem{Ccorr}D.Y. Ivanov, A. Schiller and V.G. Serbo, Phys.Lett. 
B454 (1999), 155, hep-ph/9809449.

\bibitem{JJ} John Jowett, Workshop on photon-induced collisions at the LHC,
CERN, june 2-4, 2014.

\bibitem{Fermi}E. Fermi, Z.Phys. A29, (1924), 315, doi:10.1007/BF03184853.

\bibitem{Williams}E.J. Williams, Kong.Dan.Vid.Sel.Mat.Fys.Med. 13N4 (1935) 4,1.

\bibitem{Weizs}C.F. von Weizs\"{a}cker, Z.Phys. 88 (1934) 612.

\bibitem{Budnev}V.M. Budnev, I.F. Ginzburg, G.V. Meledin, V.G. Serbo, 
Phys.Rept. 15 (1975) 181-281.  

\bibitem{Agger}C.K. Agger and A.H. Sorensen, Phys.Rev. A55 (1997) 402.   

\bibitem{bfcross}H. Meier, Z. Halabuka, K. Hencken, D. Trautmann and G. Baur,
Phys.Rev. A63 (2001) 032713, nucl-th/0008020. 

\bibitem{Serbo1} A.N. Artemyev, U.D. Jentschura, V.G. Serbo and A. Surzhykov,
Eur.Phys.J. C72 (2012), 1935, arXiv:1204.0263. 

\bibitem{Serbo2} A.N. Artemyev, V.G. Serbo and A. Surzhykov, Eur.Phys.J. 
C74 (2014), 2829, arXiv:1402.4305. 

\bibitem{Baur}G. Baur, K. Hencken, D. Trautmann, S. Sadovsky and Y. Kharlov,
Phys.Rept. 364 (2002) 359. 

\bibitem{Ginzburg}I.F. Ginzburg and A. Schiller, Phys.Rev.D60 (1999) 075016, hep-ph/9903314.

\bibitem{L3_mono}L3 Collaboration (M. Acciarri et al.), Phys.Lett.B345 (1995)
609. 

\bibitem{D0_mono}D0 Collaboration (B. Abbott et al.), Phys.Rev.Lett. 81 (1998),
524, hep-ex/9803023.


\end{thebibliography}
\end{document}